\documentclass[aps,prc,twocolumn,showpacs,preprintnumbers,amsmath,amssymb,superscriptaddress,floatfix]{revtex4-1}

\usepackage{graphicx}% Include figure files
\usepackage{dcolumn}% Align table columns on decimal point
\usepackage{bm}% bold math
\usepackage{epsfig}
\usepackage{dcolumn}% Align table columns on decimal point
\usepackage{epstopdf}
\usepackage{float}
\usepackage{textcomp}
\usepackage[sc]{mathpazo}
\usepackage{mathtools}
\usepackage[Symbol]{upgreek}
\usepackage[utf8]{inputenc}
\usepackage{float}
\usepackage{footnote}
\usepackage{titlesec}

\begin{document}
	
\title{Statistical properties of the well deformed $^{153,155}$Sm nuclei and the scissors resonance}
	
\author{K.\ L.\ Malatji}
\email{klmalatji@tlabs.ac.za}
\affiliation{Department of Subatomic Physics, iThemba LABS, P.O. Box 722, Somerset West 7129, South Africa}
\affiliation{Physics Department, Stellenbosch University, Matieland 7602, South Africa}
	
\author{K.\ S.\ Beckmann}
\email{krissbec@fys.uio.no}
\affiliation{Department of Physics, University of Oslo, N-0316, Oslo, Norway\\}%
	
\author{M.\ Wiedeking}
\email{wiedeking@tlabs.ac.za}
\affiliation{Department of Subatomic Physics, iThemba LABS, P.O. Box 722, Somerset West 7129, South Africa}
\affiliation{School of Physics, University of the Witwatersrand, Johannesburg 2050, South Africa}
	
\author{S.\ Siem}
\affiliation{Department of Physics, University of Oslo, N-0316, Oslo, Norway\\}
	
\author{S.\ Goriely}
\affiliation{Institut d'Astronomie et d'Astrophysique, Universit\'e Libre de Bruxelles, CP 226, B-1050 Brussels, Belgium\\}
	
\author{A.\ C.\ Larsen}
\affiliation{Department of Physics, University of Oslo, N-0316, Oslo, Norway\\}
	
\author{K.\ O.\ Ay}
\affiliation{Department of Physics, Faculty of Science and Letters, Eskisehir Osmangazi University, TR-26040 Eskisehir, Turkey\\}
	
\author{F.\ L.\ Bello Garrote}
\author{L. Crespo Campo}
\author{A.\ G{\"o}rgen}
\affiliation{Department of Physics, University of Oslo, N-0316, Oslo, Norway\\}

\author{M.\ Guttormsen}
\author{V.\ W.\ Ingeberg}
\affiliation{Department of Physics, University of Oslo, N-0316, Oslo, Norway\\}
	
\author{P.\ Jones}
\affiliation{Department of Subatomic Physics, iThemba LABS, P.O. Box 722, Somerset West 7129, South Africa}

\author{B.\ V.\ Kheswa}
\affiliation{Department of Subatomic Physics, iThemba LABS, P.O. Box 722, Somerset West 7129, South Africa}
\affiliation{Department of Applied Physics and Engineering Mathematics, University of Johannesburg, Doornfontein 2028, South Africa\\}
	
\author{P.\ von\ Neumann-Cosel}
\affiliation{Institut f\"{u}r Kernphysik, Technische Universit\"{a}t Darmstadt, D-64289 Darmstadt, Germany\\}
	
\author{M.\ Ozgur}
\affiliation{Department of Physics, Faculty of Science and Letters, Eskisehir Osmangazi University, TR-26040 Eskisehir, Turkey\\}
	
\author{G.\ Potel}
\affiliation{Lawrence Livermore National Laboratory, Livermore, California 94551, USA}
	
\author{L.\ Pellegri}
\affiliation{Department of Subatomic Physics, iThemba LABS, P.O. Box 722, Somerset West 7129, South Africa}
\affiliation{School of Physics, University of the Witwatersrand, Johannesburg 2050, South Africa}
	
\author{T.\ Renstr{\o}m}
\author{G.\ M.\ Tveten}
\affiliation{Department of Physics, University of Oslo, N-0316, Oslo, Norway\\}%

\author{F.\ Zeiser}
\affiliation{Department of Physics, University of Oslo, N-0316, Oslo, Norway\\}%
	
\date{\today}
	
\begin{abstract}
		
The Nuclear Level Densities (NLDs) and the $\gamma$-ray Strength Functions ($\gamma$SFs) of $^{153,155}$Sm have been extracted from (d,p$\gamma$) coincidences using the Oslo method. The experimental NLD of $^{153}$Sm is higher than the NLD of $^{155}$Sm, in accordance with microscopic calculations. The $\gamma$SFs of $^{153,155}$Sm are in fair agreement with QRPA calculations based on the D1M Gogny interaction. An enhancement is observed in the $\gamma$SF for both $^{153,155}$Sm nuclei around 3 MeV in excitation energy and is attributed to the M1 Scissors Resonance (SR). Their integrated strengths were found to be in the range 1.3\ ---\ 2.1 and 4.4\ ---\ 6.4 {$\mu^{2}_{N}$} for $^{153}$Sm and $^{155}$Sm, respectively. The strength of the SR for $^{155}$Sm is comparable to those for deformed even-even Sm isotopes from nuclear resonance fluorescence measurements, while that of $^{153}$Sm is lower than expected.

\end{abstract}
	
\maketitle
	
\section{Introduction}
\label{sec:level1}
	
The stable samarium isotopic chain provides an excellent opportunity to systematically investigate the evolution of nuclear structure effects, from the semi-magic and near spherical $^{144}$Sm to the highly-deformed $^{154}$Sm isotope. As the nuclear shape changes, statistical quantities such as the Nuclear Level Density (NLD) and $\gamma$-ray Strength Function ($\gamma$SF) are expected to change and provide evolutionary information across the isotopic chain. Furthermore, the behavior of resonance modes, such as the $M1$ Scissors Resonance (SR), $E1$ Pygmy Dipole Resonance (PDR), and the Low-Energy Enhancement (LEE), can be tracked.
	
The strength of the SR is sensitive to the ground state deformation \cite{Richter1990, Margraf1993,Goriely2016,Goriely2019a}. The SR was first predicted in 1978 by Lo Iudice and Palumbo \cite{Iudice1978} before it was observed experimentally a few years later \cite{Bohle1984}. Even-even nuclei were initially considered to be the best experimental candidates to exhibit strong SR modes. However, it soon became apparent that this mode should also present in odd-even and odd-odd systems, although its intensity may be fragmented significantly, making it more difficult to detect \cite{Enders1997,Heyde2010}. Since then, many heavy deformed even-even and odd-mass rare-earth nuclei have been systematically investigated. So far, the SR mode has been experimentally observed in vibrational and rotational \cite{Heyde2010}, as well as in $\gamma$-soft nuclei \cite{Linnemann2003,Garrel2006} and has also been observed in the actinide region \cite{Margraf1990,Guttormsen2012, Guttormsen2013, Guttormsen2014, Tornyi2014, Laplace2016, Zeiser2019}, as well as in the rare-earth mass region \cite{Voinov2001,Melby2001,Siem2002,Guttormsen2003,Krticka2004,Agvaanluvsan2004,Nyhus2010,Baramsai2015,Simon2016,Renstrom2018,Goriely2019a}. 
	
A range of different experimental techniques have been used to investigate the low-energy SR. These include ground state absorption experiments such as inelastic electron scattering \cite{Bohle1984}, nuclear resonance fluorescence (NRF) \cite{Kneissl1996} and average resonance capture (ARC) \cite{Kopecky2017}, as well as $\gamma$-decay experiments such as radiative neutron capture \cite{Krticka2004,Baramsai2015} and the Oslo method \cite{Schiller2000,Larsen2011}, which extracts information from ion scattering or transfer reactions. In general, $\gamma$-decay experiments have yielded larger SR strengths than the ground state absorption experiments. Summaries of experimental techniques to measure $\gamma$SFs can be found in Goriely {\it et al.} \cite{Goriely2019}.

%A range of different experimental techniques have been used to investigate the low-energy SR. These include inelastic electron scattering \cite{Bohle1984}, nuclear resonance fluorescence (NRF) \cite{Kneissl1996}, average resonance capture (ARC) \cite{Kopecky2017}, radiative neutron capture \cite{Krticka2004,Baramsai2015} and the Oslo method \cite{Schiller2000,Larsen2011}, which extracts information from ion scattering or transfer reactions. Summaries of experimental techniques to measure $\gamma$SFs can be found in Goriely {\it et al.} \cite{Goriely2019}.
	
Recent measurements of the $\gamma$SF in the actinides \cite{Guttormsen2012,Laplace2016, Tornyi2014} have uncovered that the SR exhibits a pronounced double-hump structure, seemingly independent of whether the nucleus has an even or odd number of neutrons. The splitting has also been observed, albeit weaker, in the transitional nucleus $^{181}$Ta \cite{Angell2016,Brits2019}. It is suggested that the splitting of the SR may be due to the isovector spin-scissors mode \cite{Balbutsev2018}, or due to triaxiality \cite{Iachello1981}. The splitting in the SR has not been reported for any of the rare-earth nuclei studied with the NRF technique \cite{Pietralla1998}, with the Oslo method \cite{Voinov2001,Melby2001,Siem2002,Guttormsen2003,Agvaanluvsan2004,Nyhus2010,Simon2016,Renstrom2018} or other techniques \cite{Krticka2004,Baramsai2015,Goriely2019a}.
	
The $\gamma$SFs of the isotopes $^{148,149}$Sm were measured, and analyzed with the Oslo method, almost two decades ago and a weak structure was identified to possibly be due to the SR, called a pygmy at the time \cite{Siem2002}. With data already available on these two weakly deformed isotopes, together with the recent measurements of $^{151,153}$Sm \cite{Simon2016} (also analyzed with the Oslo method), it is interesting to extend the investigation towards more deformed Sm nuclei where the SR is expected to be more prominent. 
	
In this paper, the measurement of the NLDs and $\gamma$SFs for the odd-even $^{153,155}$Sm extracted with the Oslo method from the $^{152,154}$Sm(d,p$\gamma$) reactions are reported. The integrated strength of the SR in both isotopes are extracted and compared to previous experimental data. In Sec.\ \ref{sec2} the experimental setup is presented and Sec.\ \ref{sec3} provides a brief overview of the Oslo method and the normalization procedures used, as well as the measured NLDs and $\gamma$SFs. Section\ \ref{sec6} investigates the presence of the SR in Sm isotopes and its integrated strength. Discussion follows in Sec.\ \ref{sec7} and a brief summary in Sec.\ \ref{sec8}.

\section{\label{sec2}Experimental Setup}

Two experiments were performed at the Oslo Cyclotron Laboratory (OCL) at the University of Oslo using self-supporting $^{152}$Sm (enriched to 98.3\%) and $^{154}$Sm (enriched to 98.7$\%$) targets with thicknesses of 2.9 and 3.2 mg/cm$^2$, respectively. Deuteron beams of 13.5 MeV and 13 MeV were used to populate excited states in $^{153,155}$Sm. The SiRi particle telescope \cite{Guttormsen2011} and CACTUS scintillator \cite{Guttormsen1990} arrays were used to detect charged particles and $\gamma$-rays in coincidence.
	
The $\Delta E$-$E$ SiRi particle-telescope consisted of eight 130~$\mu$m thin, segmented silicon $\Delta E$ detectors and eight 1550~$\mu$m thick $E$ silicon detectors. These were placed at backward angles to reduce detection of elastically scattered events and covered a polar angular range of $\theta_{lab} = 126^{\circ} - 140^{\circ}$ with respect to the beam direction. The energy resolution, as determined from the elastic peaks, is $\sim$ 130 keV. The CACTUS array consisted of 26 and 24 NaI(Tl) detectors for the $^{153}$Sm and $^{155}$Sm measurements, respectively. The $5" \times 5"$ crystals were positioned 22 cm from the target, covering solid angles of 17$\%$ and 15$\%$ of $4\pi$ sr, respectively. CACTUS has an energy resolution of 7$\%$ FWHM for a 1.332 MeV $\gamma$-ray transition.

\begin{figure*}
\centering
\includegraphics[scale=0.88]{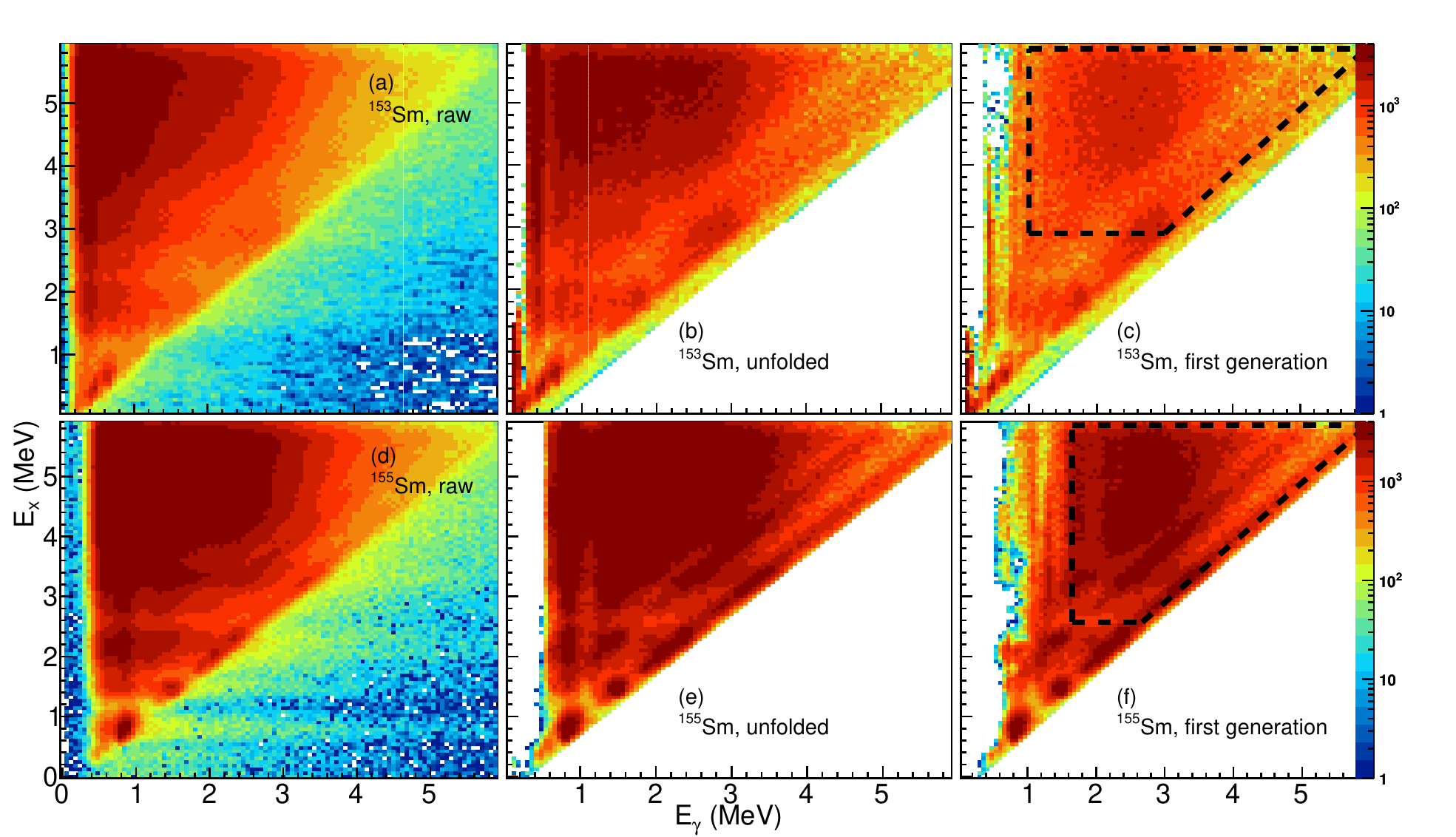}
\caption{(Color online) The raw (panel a, d), unfolded (panel b, e) and first generation (primary) $\gamma$-ray matrix (panel c, f) of $^{153,155}$Sm. The dashed lines indicate the energy regions from which the NLDs and $\gamma$SFs were extracted.}
\label{alfnas}
\end{figure*}

The $E$ detectors provided a start signal and the NaI(Tl) detectors a stop signal for the time-to-digital converters, enabling event-by-event sorting for the $\gamma$-particle coincidence data. Calibration of the SiRi and CACTUS detectors was achieved using distinct $\gamma$-ray transitions of $^{29}$Si obtained from $^{28}$Si(d,p$\gamma$) calibration runs which provided well-resolved particle and $\gamma$-ray peaks. During offline analysis, charged-particle-$\gamma$ coincidence events were extracted within a prompt time gate of 20 ns. Equivalently wide time gates were used to remove the majority of randomly correlated events from the prompt particle-$\gamma$ events. The excitation energy ($E_{x}$) versus $\gamma$-ray energy ($E_{\gamma}$) matrices were constructed from the particle-$\gamma$ coincidence events and are shown in Fig.\ \ref{alfnas} (a) and (d). The Oslo method \cite{Schiller2000,Larsen2011} is applied on these matrices to extract simultaneously the NLDs and the $\gamma$SFs up to the neutron-binding energies ($B_n$) through several iterative methods, discussed in the next section. The analysis in this work was performed using the Oslo method software version 1.1.2 \cite{Guttormsen2018}.

\section{\label{sec3}Data Analysis}
\subsection{\label{sec3a}\textbf{The Oslo Method}}

The $\gamma$-ray spectra, extracted for each $E_{x}$ bin, were unfolded with the iterative procedure of Ref.\ \cite{Guttormsen1996} and then corrected for the known NaI(Tl) response functions remeasured in 2012 \cite{Campo2016}, in order to obtain the full-energy $\gamma$-ray spectra. At this point, the first-generation $\gamma$-ray method \cite{Guttormsen1987} is used to extract the primary $\gamma$-rays from the decay cascades in each $E_{x}$ bin. The raw matrices (a) and (d), the unfolded matrices (b) and (e), and the resulting first generation $\gamma$-ray matrices (c) and (f), P($E_{\gamma}$, $E_{x}$), are summarized in Fig.\ \ref{alfnas}. The diagonals where $E_{\gamma}$=$E_{x}$ represent all direct decays to the ground state. 
	
For $^{153}$Sm, the region used for extraction of the NLD and $\gamma$SF was from $E_{\gamma}$ = 0.96 MeV to and including the $E_{x}$ = $E_{\gamma}$ diagonal, and from $E_{x}$ = 2.88 MeV up to $E_{x}$ = 5.69 MeV. For $^{155}$Sm, the limits were from $E_{\gamma}$ = 1.65 MeV to and including the $E_{x}$ = $E_{\gamma}$ diagonal, and from $E_{x}$ = 2.49 MeV up to $E_{x}$ = 5.73 MeV. They were chosen to exclude regions characterized by discrete transitions at low excitation energies. The regions in Fig.\ \ref{alfnas} (c) and (f) that correspond to $E_{\gamma}<$ 1 MeV have low statistics due to over-subtraction of discrete and strong $\gamma$-ray transitions. This energy region is therefore also excluded from further analysis. 

The NLDs and $\gamma$SFs of $^{153,155}$Sm were extracted simultaneously from the $P$($E_{\gamma}$,$E_{x}$) matrix, using the ansatz \cite{Tveter1996,Midtboe2019}:\begin{equation}
\centering
\label{eq1}
P(E_{x},E_{\gamma}) \propto \rho(E_{x}-E_{\gamma})\mathcal{T}(E_{\gamma}),
\end{equation}
where $\rho(E_{x}-E_{\gamma})$ is the level density at the final $E_{x}$ to which the nucleus decays. The parameter $\mathcal{T}(E_{\gamma})$ is the $\gamma$-ray transmission coefficient, and assuming the generalized Brink-Axel Hypothesis \cite{Brink1957,Axel1962}, it is only dependent on the $\gamma$-ray energy. %and $\mathcal{T}(E_{\gamma})$ is the $\gamma$-ray transmission coefficient, which only depends on the $\gamma$-ray energy in accordance with the assumption of the generalized Brink-Axel Hypothesis \cite{Brink1957,Axel1962}.
It is also assumed in Eq.\ (\ref{eq1}) that the $\gamma$-decay pattern from any initial excitation energy is independent of whether the nucleus was populated into this excitation energy directly from a nuclear reaction or by $\gamma$-ray decays from higher-lying states \cite{Bohr1969,Henden1995}. A $\chi^2$ minimization is performed, between the experimental $P$($E_{\gamma}$,$E_{x}$) and a theoretical $P_{theo}$($E_{\gamma}$,$E_{x}$) in which $\rho(E_{x}-E_{\gamma})$ and $\mathcal{T}(E_{\gamma})$ are treated as free parameters \cite{Schiller2000}:
 
\begin{equation}
\centering
\begin{split}
\tilde{\rho}(E_{x}-E_{\gamma}) &= A \rho(E_{x}-E_{\gamma})\exp[{\alpha (E_{x}-E_{\gamma}})],\\
\tilde{\mathcal{T}}(E_{\gamma})&= B \mathcal{T}(E_{\gamma})\exp(\alpha E_{\gamma}).
\end{split}
\label{q2}
\end{equation}
 
The transformation parameters $\alpha$, $A$ and $B$ correspond to physical solutions and are deduced from external experimental data to get the solution to Eq. (\ref{eq1}). At this point, the features in the NLDs and $\gamma$-ray transmission coefficients are fixed, except for the slopes and absolute values. Note that the resulting $\rho$ and $\mathcal{T}$ functions do not depend on the initial parameters used in the iterative procedure.
	
\subsection{\label{sec4}Normalization of the NLDs}
	
\begin{figure}
\centering
\includegraphics[scale=0.56]{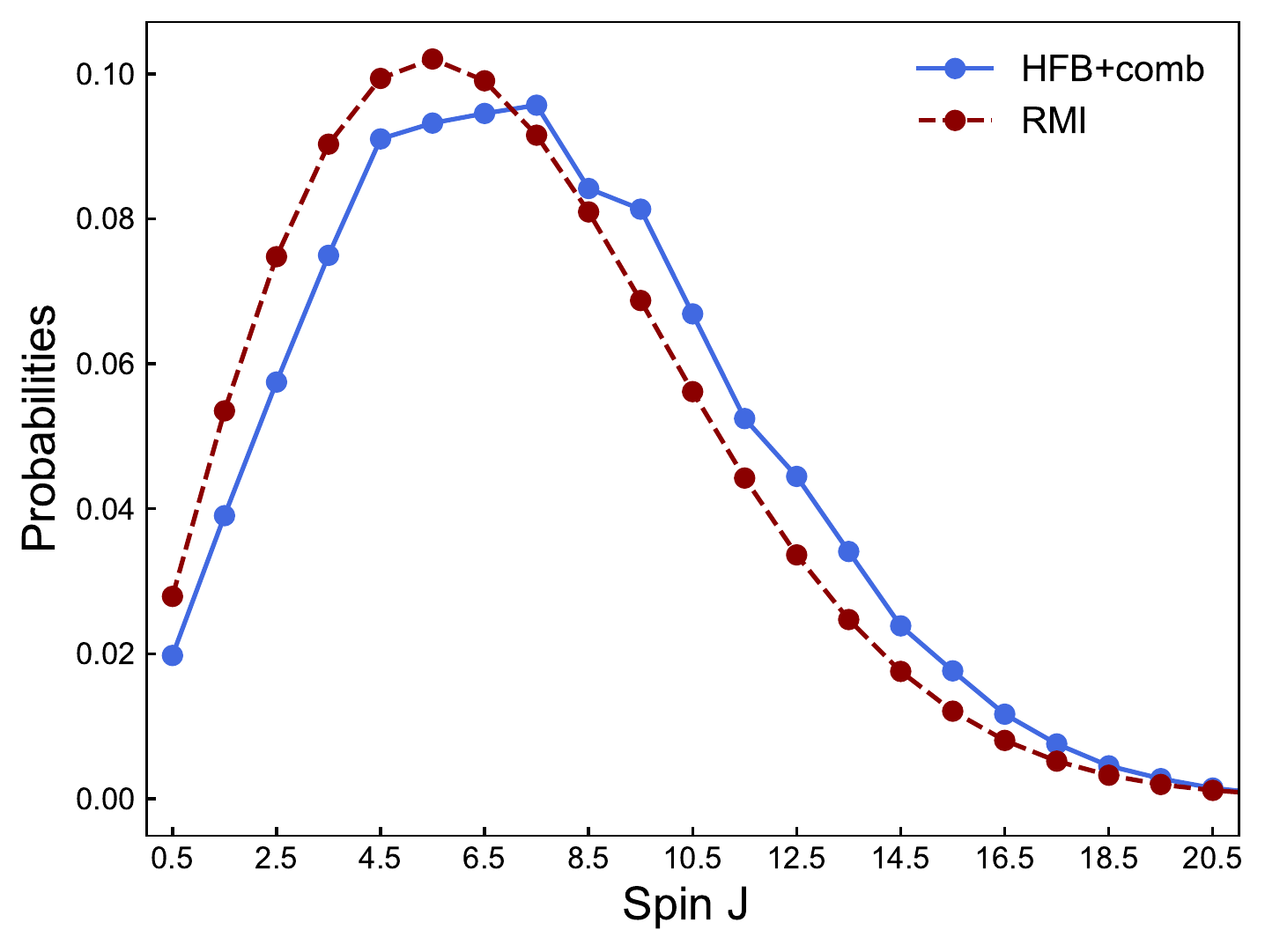}
\caption{(Color online) Spin distributions at $B_n$ estimated from HFB+comb (blue solid line) and RMI (red dashed line) models for $^{153}$Sm. }
\label{155Sm_spin}
\end{figure}
	
The extracted NLD is normalized to the known experimental discrete states \cite{NNDC} at $E_{x}\lesssim$ 1 MeV and extrapolated to the NLD, $\rho(B_n)$, at the neutron-binding energy ($B_n$), determining its slope and absolute value. The level density $\rho(B_n)$ is determined from the average $s-$wave neutron-resonance spacing $D_{0}$ \cite{Capote2009a}, using Eq.~(28) of Ref.\ \cite{Schiller2000}. The parameters used for the normalization are listed in Tab.\ \ref{t1x}. Due to the unavailability of experimental data on the spin and parity $J^{\pi}$ distribution at $B_n$, the Rigid Moment of Inertia formula (RMI) which assumes equiparity \cite{Egidy2006} and the Hartree-Fock-Bogoliubov plus Combinatorial (HFB+comb) \cite{Goriely2008} models were utilized to model the distributions. The RMI and HFB+comb spin distributions at ${B_n}$ are shown in Fig.\ \ref{155Sm_spin} for $^{153}$Sm.
\begin{figure}
\centering
\includegraphics[scale=0.7]{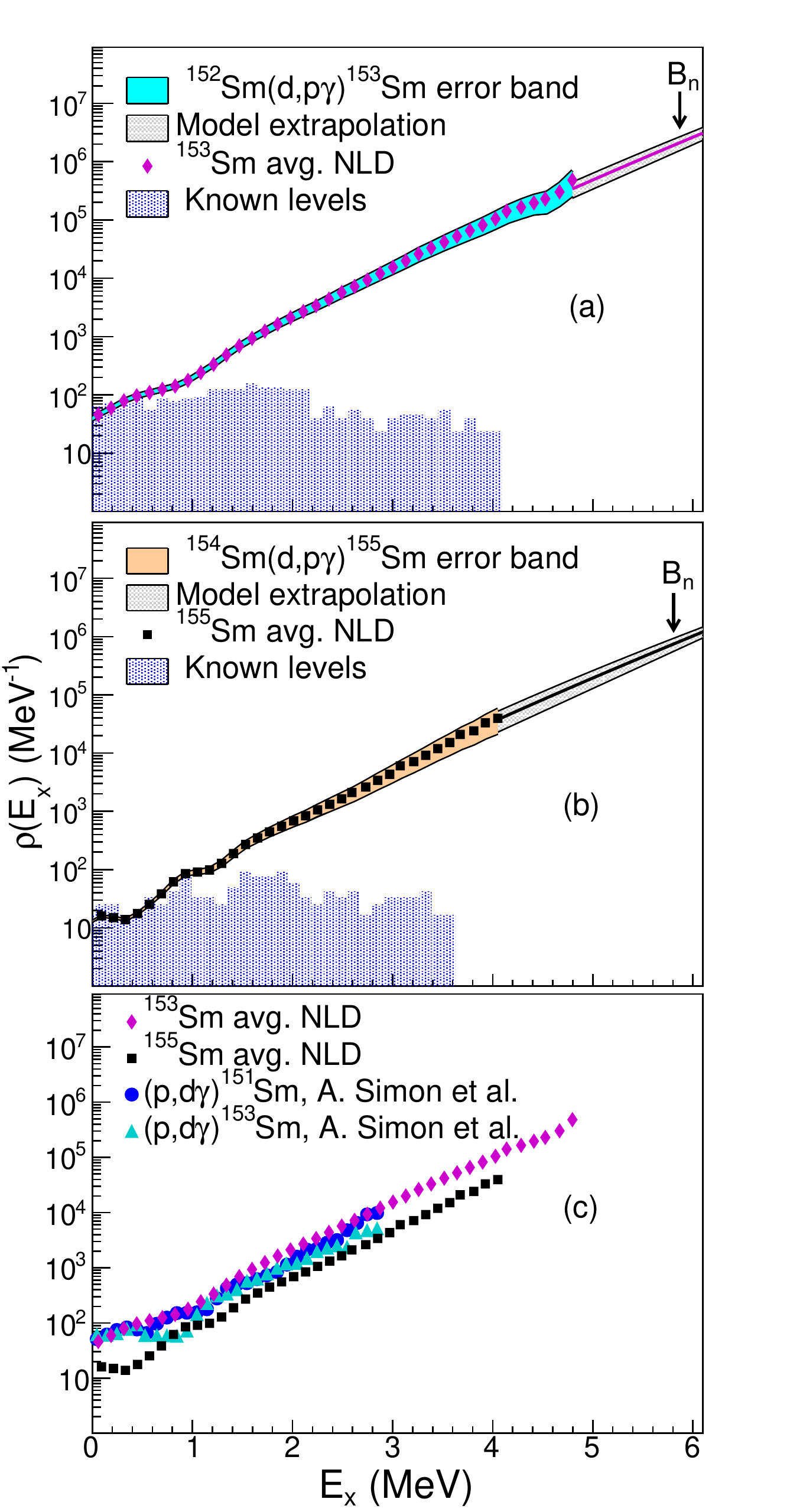}
\caption{(Color online) Experimental NLDs of (a) $^{153}$Sm and (b) $^{155}$Sm from the present (d,p$\gamma$) experiments, normalized using the RMI and HFB+comb models and extrapolated with the CT and Fermi Gas models to match the estimated $\rho$(B$_n$). The error bars on the NLD data points represent only statistical uncertainties, whereas the error band represent both statistical uncertainties as well as systematic uncertainties (see text for details). The error bands beyond the last data point (grey shaded) indicate the range of the extrapolation models and the solid line represent the average of the limits. The NLDs of $^{153,155}$Sm are compared to the NLDs of (p,d$\gamma)^{151,153}$Sm from Ref.\ \cite{Simon2016} in panel (c).}
\label{fig:NLDs}
\end{figure}

The HFB+comb model is a global microscopic approach to calculate the energy- and $J^{\pi}$- dependent NLD \cite{Goriely2008}. The HFB+comb model can be renormalized to match the known experimental discrete states and the average $s-$wave neutron-resonance spacing $D_{0}$, as detailed in Ref.~\cite{Goriely2008}.
	
Figures\ \ref{fig:NLDs} (a) and (b) present the extracted NLDs for $^{153,155}$Sm. At low $E_{x}$ the NLDs follow closely the experimental discrete states \cite{NNDC}. The NLD-bands represent the standard deviation of the level densities as a result of the statistical errors stemming from the Oslo method \cite{Schiller2000,Larsen2011}, as well as systematic errors. The systematic errors take into account the variations of slope and absolute value due to normalizing using the HFB+comb (upper limit) and RMI (lower limit) as well as from varying $D_{0}$ within its uncertainties and varying a reduction factor which is used to scale the width of the spin distribution at $B_n$. The data points in Fig.\ \ref{fig:NLDs} (a) and (b) represent the average of the upper and lower limits with the statistical error bars. It is important to distinguish between statistical and systematic errors because the statistical errors are small and they limit the possibility of fluctuations between neighboring points. As can be seen in Fig.\ \ref{fig:NLDs}, the extracted experimental NLD is not available up to $B_n$ due to the exclusion of transitions below the E$_\gamma$-cutoff discussed in Sec.\ \ref{sec3a}. In order to bridge the gap between the highest $E_{x}$ data point and $\rho(B_n)$, an extrapolation is performed using the Fermi Gas (upper limit) \cite{Egidy1988,Egidy2006} and Constant Temperature (lower limit) \cite{Ericson1959,Gilbert1965} models.
	
In Fig.\ \ref{fig:NLDs} (c) the NLDs of $^{153,155}$Sm are compared to the NLDs of (p,d$\gamma)^{151,153}$Sm from Ref.\ \cite{Simon2016} and are shown to be in reasonable agreement. Note that the level density is decreasing significantly from $^{153}$Sm to $^{155}$Sm. This will be discussed in Sec. \ref{sec7}.
	
\begin{table*}
\centering
\caption{Parameters used for extraction and normalization of $\rho(E_{x})$ and $\mathcal{T}(E_\gamma)$ in $^{153,155}$Sm. The temperature parameter $T_{CT}$ is used for the CT model extrapolation to the $B_n$.}
\setlength{\tabcolsep}{5.5pt}
\begin{tabular}{lccccccccccc}
\hline
\hline
Isotope&$J_{\pi}$&$B_n$& $a$ & $E1$ & $T_{CT}$ & $\sigma(B_n)_{RMI}$&$D_0$&$\rho_{RMI}(B_n)$&$\rho_{HFB}(B_n)$&$\langle \Gamma_{\gamma}(B_n)\rangle$\\
~&~&(MeV)&(MeV$^{-1}$)&(MeV)&(MeV)&~&(eV)&($10^{6}$MeV$^{-1}$)&($10^{6}$MeV$^{-1}$)&(meV)\\
\hline
\\
$^{153}$Sm&$3/2^{+}$&5.868&18.5&-0.66&0.57&6.0(6)$^{a}$&41.0(28)$^{b}$&1.77(36)$^{a}$&2.52(49)&60.0$^{c}$(37)$^{d}$(110)$^{e}$\\ 
$^{155}$Sm&$3/2^{-}$&5.807&18.0&-0.56&0.55&6.1(6)$^{a}$&112(15)$^{b}$&0.66(16)$^{a}$&0.80(17)&74$^{c}$(11)$^{d}$(13)$^{e}$\\ 
\hline
\end{tabular}
\\
\begin{flushleft}
$^{a}${Calculated with the rigid moment of inertia formula of von Egidy and Bucurescu \cite{Egidy2006}.}\\
$^{b}${Taken from Ref.\ \cite{Mughabghab2018}.}\\
$^{c}${Weighted average, $\bar{x}$, calculated from the $x_i\pm\sigma_i$ resonances listed in Ref.\ \cite{Mughabghab2018}, using $\bar{x}= \sum_{i=1}^Nw_ix_i/\sum_{i=1}^Nw_i$, where $w_i={1}/{\sigma_i^2}$}.\\
$^{d}${Uncertainty propagation calculated using $\sigma_{\bar{x}}= \sqrt{1/\sum_{i=1}^N{w_i}}$.}\\
$^{e}${Standard deviation of the weighted average calculated using $\sigma_{\bar{x},std}= \sqrt{N'\sum_{i=1}^Nw_i(x_i-\bar{x})^2/(N'-1)\sum_{i=1}^Nw_i}$, where $N'$ is the number of non-zero weights.}\\
\label{t1x}
\end{flushleft}
\end{table*}
	
The (d,p$\gamma$) reaction may populate a limited spin range due to its low-angular momentum transfer. This again influences the primary $\gamma$-ray spectra $P(E_{x}, E_{\gamma})$. A slope correction of the $\gamma$SF might therefore be necessary in particular for sub-Coulomb barrier reactions \cite{Guttormsen2012, Guttormsen2013, Guttormsen2014, Tornyi2014, Laplace2016, Zeiser2019,Ingeberg2020}. To verify whether such a correction is necessary, the $J^{\pi}$ distribution populated by the (d,p$\gamma$) reaction has been investigated with the statistical nuclear reaction code TALYS (v1.95) \cite{Koning2012} for the deuteron absorption compound reaction formation assuming isotropic emission. The non-elastic breakup, in which the neutron is absorbed by the target in a two-step (direct deuteron breakup + neutron absorption) mechanism has been investigated, as a function of the excitation energy of the residual nucleus following the Green's function transfer formalism of Refs. \cite{Potel2015,Potel2017}. The results indicate low non-elastic breakup cross sections ($\approx$ 2\ -\ 3 mb/sr MeV) for $\theta_{lab} = 126^{\circ} - 140^{\circ}$, and the deuteron-fusion proton-evaporation dominates, leading to a broad spin distribution. Therefore, the discrepancy in the populated spin range is considered small and no slope correction was performed.

\begin{figure*}
\centering
\includegraphics[scale=0.74]{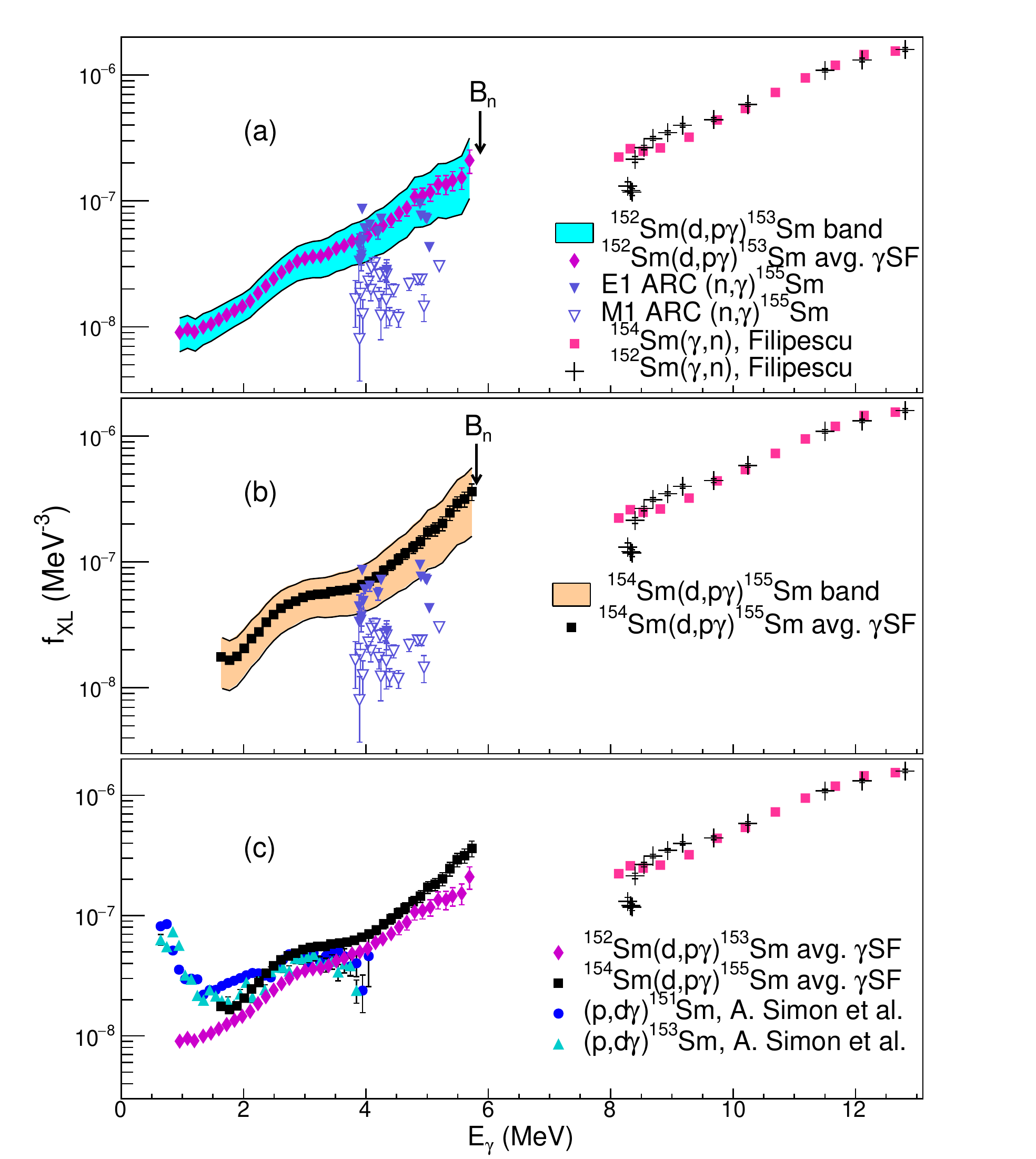}
\caption{(Color online) The experimental $\gamma$SF of $^{153}$Sm (a) and $^{155}$Sm (b) compared to the $^{152,154}$Sm photo-nuclear reaction GEDR measurements of \cite{Filipescu2014}, and $E$1 and $M$1 (n,$\gamma$)$^{155}$Sm data measured in ARC experiments \cite{Kopecky2017}. The $\gamma$SFs of $^{153}$Sm and $^{155}$Sm, which only include the systematic errors, are compared in panel (c). The $\gamma$SFs of $^{151,153}$Sm from Ref.\ \cite{Simon2016} are included for comparison.}
\label{GSF}
\end{figure*}

\subsection{\label{sec5}Normalization of the $\gamma$SFs}

The $\gamma$-ray transmission coefficient, $\mathcal{T}_{XL}(E_{\gamma})$, of multipolarity $L$ and electromagnetic character $X$ (electric, $E$, or magnetic, $M$) is transformed to the total experimental $\gamma$SF through the relationship \cite{Capote2009a}: \begin{equation}
\centering
\begin{split}
f(E_{\gamma}) &\approx f_{E1}(E_{\gamma}) + f_{M1}(E_{\gamma})\\&\approx\frac{1}{2\pi E_{\gamma}^{3}}B [\mathcal{T}_{E1}(E_\gamma) + \mathcal{T}_{M1}(E_\gamma)],
\end{split}
\label{transcoff}
\end{equation} 
	
\noindent assuming dominance of dipole transitions for statistical $\gamma$-ray decays. This assumption is strongly supported by data, see e.g.\ Ref.\ \cite{Kopecky1987}. The absolute normalization parameter $B$ in Eq.\ (\ref{transcoff}) is determined using the experimental average total radiative width $\langle \varGamma_{\gamma} \rangle_{\ell=0}$ at $B_n$. The corresponding parameters obtained and used for the normalization are summarized in Tab.~\ref{t1x}. The value of $\langle \varGamma_{\gamma}(B_n) \rangle_{\ell=0}$ was obtained by calculating the weighted average of the resonance widths listed in Ref.~\cite{Mughabghab2018}. 
 
Using the total average radiative width, the parameter $B$ is determined by \cite{Kopecky1990,Schiller2000}:
\begin{equation} \label{eq: Beq}
\centering
\begin{split}
\langle{\Gamma_{\gamma}(B_n)}\rangle=\frac{D_0B}{2\pi}&\int_0^{B_n}dE_{\gamma}\mathcal{T}(E_{\gamma})\times \rho(B_n-E_{\gamma})\cdot\\&\sum_{J=-1}^{J=1}g(B_n- E_{\gamma},J_t+J\pm1/2), 
\end{split}
\end{equation}
	
\noindent where $J_t$ is the target spin (0 for $^{152,154}$Sm) and the function $g(E_{x},J)$ is the relative probability of a given spin at excitation energy $E_{x}$. In order to calculate the integral in Eq.\ (\ref{eq: Beq}), a log-linear function for the $\gamma$SF is fitted for $\gamma$-energies between 0 and $E_{\gamma,l}$ and from $E_{\gamma,h}\rightarrow B_{n}$, where $E_{\gamma,l}$ and $E_{\gamma,h}$ represent $\gamma$-ray energies for the lowest and highest data point, respectively. For the RMI, a function approximating the spin distribution is implemented \cite{Bethe1936,Ericson1959}:
	
\begin{align} \label{eq: SpinG}
g(\sigma(E_x),J)=\frac{1}{2\sigma(E_x)^2}(2J+1)\text{exp}\left[\frac{-(J+\frac{1}{2})^2}{2\sigma(E_x)^2}\right],
\end{align}
with
\begin{align}
\sigma^{2}(E_x)=0.0146A^{5/3}\frac{1+\sqrt{1+4a(E_{x}-E_{1})}}{2a},
\label{EB6}
\end{align}

\noindent where $a$ is the level density parameter, $E_1$ is the total back-shift parameter and the spin-cutoff parameter, $\sigma(E_x)$, is a modeled variable related to the width of the distribution. However, in the HFB+comb model the explicit probability for each spin is given, as illustrated in Fig.\ \ref{155Sm_spin}. Therefore, the relevant spin probabilities required for Eq.\ (\ref{eq: Beq}) are directly obtained from the tables \cite{Koning2012}, while correcting for the excitation-energy shift used to normalize the NLDs.

The extracted experimental $\gamma$SFs of $^{153,155}$Sm are shown in Fig.\ \ref{GSF}. As in the case of the NLDs, the $\gamma$SF-bands include both statistical and systematic errors. Here, the systematic errors also take into account the uncertainty of the $\langle \varGamma_{\gamma} \rangle$ parameter. The average of the limits and the statistical errors are shown as data points within the error bands. In Fig.~\ref{GSF}, the $^{153,155}$Sm experimental $\gamma$SFs are compared to the experimental Giant Electric Dipole Resonance (GEDR) data from $^{152,154}$Sm($\gamma$,n) photo-nuclear ($\gamma$-absorption) data \cite{Filipescu2014}, as there is no GEDR data on $^{153,155}$Sm. The $\gamma$SFs from this work appear steep in comparison to the available GEDR data, possible explanations are discussed in Sec.\ \ref{sec7}. The present data are further compared to $E1$ and $M1$ (n,$\gamma$)$^{155}$Sm data measured in average resonance capture (ARC) experiments \cite{Kopecky2017}. The cross sections $\sigma_{\gamma}(E_\gamma)$ are transformed to $\gamma$SFs using the relation \cite{Axel1962}:
\begin{equation}
\centering
f(E_{\gamma}) = \frac{1}{3\pi^{2}\hbar^{2}c^{2}}{\frac{\sigma_{\gamma}(E_\gamma)}{E_{\gamma}}},
\label{q12}
\end{equation}
where the factor $1/{3 \pi^{2} \hbar^{2} c^{2}}$ = 8.674$\times10^{-8}~{\rm mb^{-1}MeV^{-2}}$.	

Finally, in Fig.\ \ref{GSF} (c), the $\gamma$SFs of $^{153}$Sm and $^{155}$Sm are compared to each other as well as to the $\gamma$SFs of $^{151,153}$Sm from Ref.\ \cite{Simon2016}. Several points of interest emerge from this comparison, such as the absolute value difference between the $^{153}$Sm and $^{155}$Sm from the current experiment. Previous measurements on close lying nuclei give consistently similar absolute values for the $\gamma$SFs. This apparent deviation is discussed in Sec.\ \ref{sec7}. Comparing the results for $^{153}$Sm from Ref.\ \cite{Simon2016} and from the present work there are clear differences. Given that both data sets were analyzed using the Oslo method it is important to understand the differing features. Firstly, there seems to be a large discrepancy at $\gamma$-energies below 2 MeV, where the (p,d$\gamma)^{153}$Sm data indicate a strong LEE, while the (d,p$\gamma)^{153}$Sm keeps trending downward. Secondly, the SR appears to be significantly more pronounced in the (p,d$\gamma)^{153}$Sm data. This will be explored below after the $B_{SR}(M1)$ strengths are extracted and compared.

\section{\label{sec6}The Scissors Resonance}

Experiments using the Oslo method can only extract the SR built on excited states in the quasi-continuum, whereas NRF measurements can only extract SR built on the ground state. From the $\gamma$SF, the integrated reduced transition strength for magnetic dipole transitions, $B_{SR}(M1)$, is obtained by determining the shape of the resonance and numerically integrating over the distribution using the Standard Lorentzian Function (SLo) \cite{Brink1957,Axel1962,Capote2009a}, in the energy range relevant to the SR: \begin{equation}
B_{SR}=\frac{27(\hbar c)^{3}}{16 \pi}\int f_{SR}^{SLo}(E_{\gamma})dE_{\gamma},
\label{B_sr1}
\end{equation} 
\noindent where the factor $27(\hbar c)^{3}/{16 \pi}=2.5980\times10^{8}~{\mu_{N}^{2}}$MeV$^{2}$.

\begin{figure}
\centering
\includegraphics[scale=0.67]{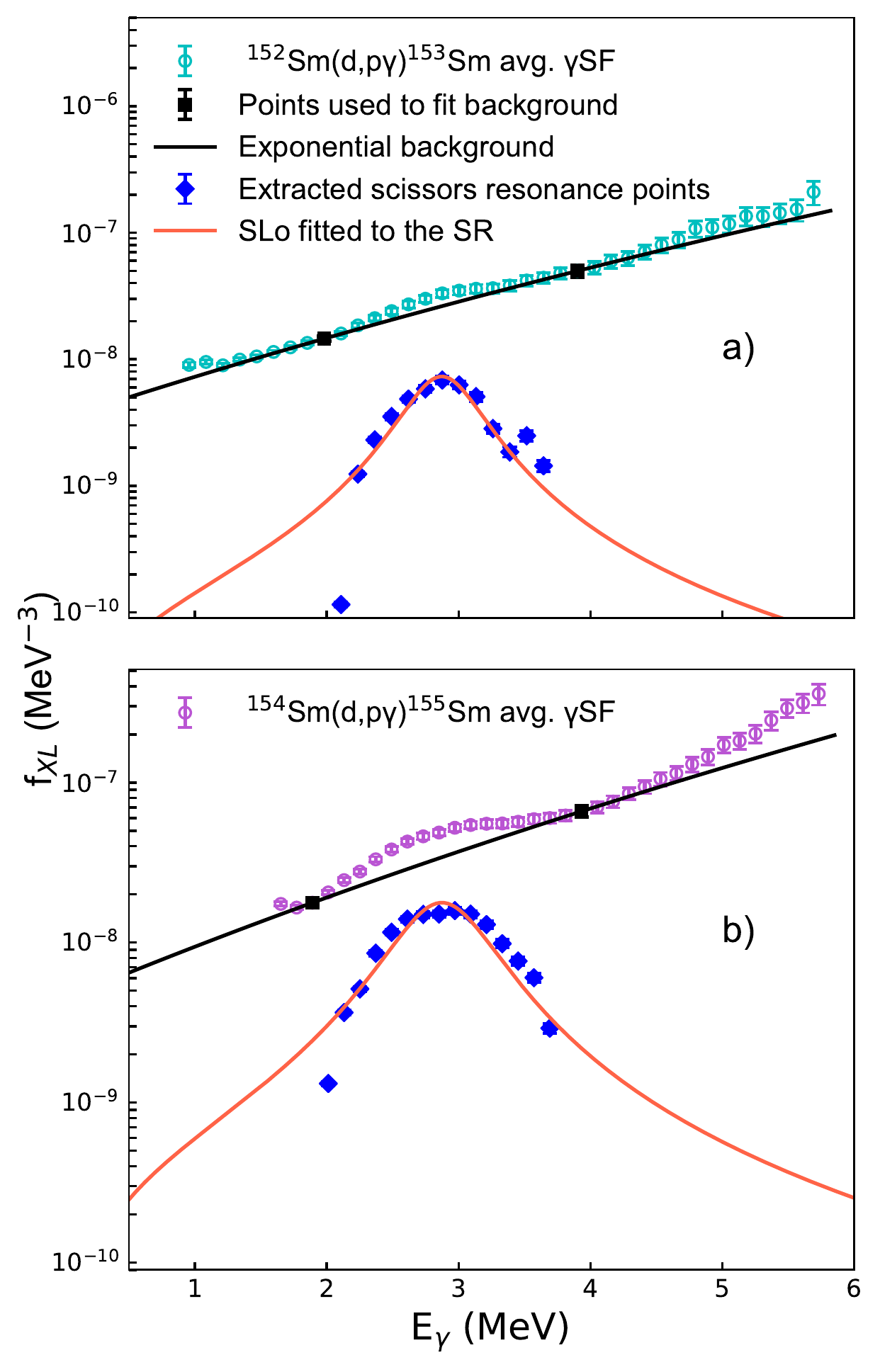}
\caption{(Color online) The extracted SR of $^{153}$Sm (a) and $^{155}$Sm (b), fitted with the exponential background and SLo (see text for details).}
\label{155_3Sm_SR}
\end{figure}

\begin{table*}
\caption{Experimental integrated $B_{SR}(M1)$ strength for samarium isotopes extracted in the given energy ranges. The quadrupole deformation of the nuclear ground state $\delta^{2}$ is taken from the FRDM12 database \cite{Moller2016}. The quantity $\omega_{SR}$ denotes the energy centroid of the SR.}
%\centering
\setlength{\tabcolsep}{13.2pt}
\begin{tabular}{lcccccccccccc}\\
			
\toprule
Isotope & Deformation & Energy range & $\omega_{SR}$ & $B_{SR}(M1)$ & Ref. & Technique
\\ 
$^{A}$X & $\delta^{2}$ & [{MeV}] & [{MeV}] & [{$\mu^{2}_{N}$}] & ~ & ~\\
\hline
\\
$^{153}$Sm& 0.26 & Full Range$^{a}$ & 2.89(0) & 1.27~---~2.13 & Present Exp.& Oslo method\\
$^{153}$Sm& 0.26 & 2.0 --- 4.0 & 2.89(0) & 1.09~---~1.85 & Present Exp.& Oslo method\\
$^{155}$Sm& 0.27 & Full Range$^{a}$ & 2.98(8) & 4.40~---~6.44 & Present Exp. &Oslo method\\
$^{155}$Sm& 0.27 & 2.0 --- 4.0 & 2.98(8) & 3.58~---~5.30& Present Exp. &Oslo method\\
\hline
\\
$^{144}$Sm&0.08&2.0 --- 4.0&3.97(4)&0.28(0)&Ziegler {\it et al.} \cite{Ziegler1990}&NRF\\
$^{148}$Sm&0.18&2.0 --- 4.0&3.07(3)&0.51(1)&Ziegler {\it et al.} \cite{Ziegler1990}&NRF\\
$^{150}$Sm&0.21&2.0 --- 4.0&3.13(3)&0.97(10)&Ziegler {\it et al.} \cite{Ziegler1990}&NRF\\
$^{151}$Sm&0.22&0.0 --- 5.0&3.00(2)&7.80(340)&Simon {\it et al.} \cite{Simon2016}&Oslo method\\
$^{152}$Sm&0.24&2.0 --- 4.0&2.99(3)&2.35(20)&Ziegler {\it et al.} \cite{Ziegler1990}&NRF\\
$^{153}$Sm&0.26&0.0 --- 5.0&3.00(2)&7.80(200)&Simon {\it et al.} \cite{Simon2016}&Oslo method\\
$^{154}$Sm&0.27&2.0 --- 4.0&3.20(3)&2.65(30)&Ziegler {\it et al.} \cite{Ziegler1990}&NRF\\
\hline
\end{tabular}
\begin{flushleft}
$^{a}${Equation \ref{B_sr1} was integrated from 0 to 20 MeV. This equates to an unrestricted range, as including higher energies did not alter the tabulated value to a significant digit.} \\
\end{flushleft}
\label{ResPar} 
\end{table*}

Several fitting methods were explored, such as including the GEDR data and fitting a collection of SLo peaks, or making a model of an exponential background plus one SLo-peak. However, none of the methods converged successfully and required many of the parameters to be fixed. Therefore a less sophisticated, but transparent method was chosen. To extract the strength of the SR an exponential function was fitted by two points in the experimental $\gamma$SFs, to approximate the background in the vicinity of the resonance, as shown in Fig.\ \ref{155_3Sm_SR}. It is deemed reasonable that the background is of exponential shape, as all conventional empirical models for the GEDR are of Lorentzian type, where the tail can be approximated by an exponential for a small energy interval. This background was then extracted from the data. An SLo was subsequently fitted to the extracted points, with statistical errors as weights in the fit. A fit was performed on the upper and lower limit of the $\gamma$SF, as well as for the average as shown in Fig.\ \ref{155_3Sm_SR}. The integrated $B_{SR}(M1)$ was found numerically from Eq.\ (\ref{B_sr1}), and the resulting strengths are listed in Tab.\ \ref{ResPar}. Due to the extraction method, a range of the SR strength is given instead of a recommended value with uncertainties. The energy centroid is consistent across the different fits and is also given with its small uncertainties in Tab.\ \ref{ResPar}.

\section{\label{sec7}Discussion}

\begin{figure}
\centering
\includegraphics[scale=0.41]{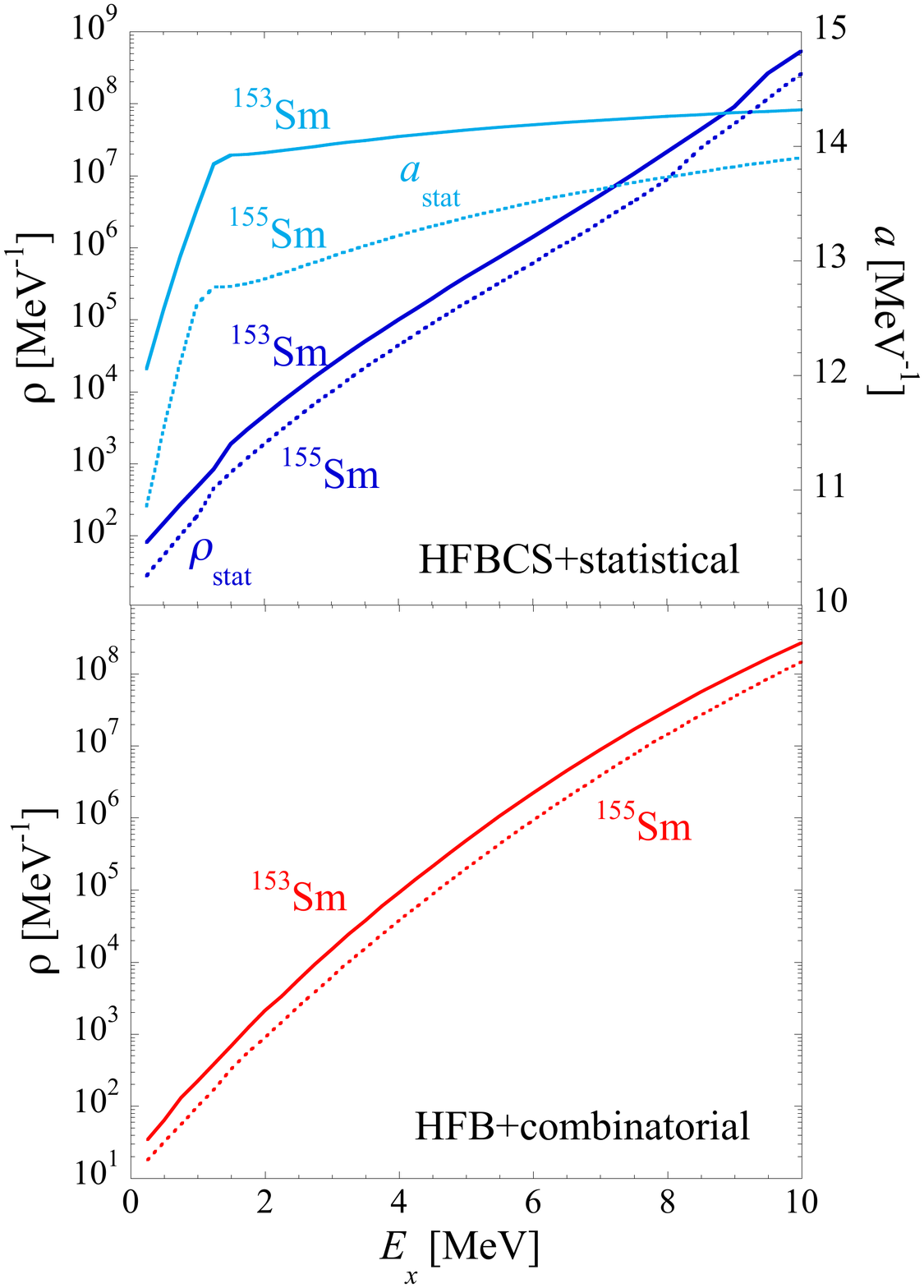}
\caption{(Color online)~Comparison between $^{153}$Sm (solid lines) and $^{155}$Sm (dashed lines) NLDs obtained within the combinatorial (red curves) \cite{Goriely2008} and statistical, $\rho_{stat}$, (dark blue curves) \cite{Demetriou2001} models. Also shown are the energy-dependent level density parameters $a_{stat}$ (light blue curves) obtained within the microscopic statistical approach \cite{Demetriou2001}.}
\label{NLD_talys}
\end{figure}

In the current experiments, the NLD and $\gamma$SF of the deformed even-odd $^{155}$Sm below $B_n$, were measured for the first time, as well as that of $^{153}$Sm in the energy range 4~$\leq E_{\gamma}\leq B_n$ MeV. 

It is interesting to note in Fig.~\ref{fig:NLDs} (c) that even though the $^{153,155}$Sm NLDs have similar slopes, the $^{153}$Sm level density is higher than that of $^{155}$Sm, which is counter-intuitive. As more neutrons are added and deformation increases, it may be expected that there should be more levels in $^{155}$Sm compared to $^{153}$Sm. The same behavior has also been observed in neodymium isotopes \cite{Ay2016,Magne_pc_2019} where the lighter, spherical or less deformed isotopes exhibit higher NLDs than the more deformed, heavier isotopes. This phenomenon is consistent with the microscopic description of the NLDs obtained within the combinatorial (HFB+combinatorial) \cite{Goriely2008} and statistical (HFBCS+statistical) \cite{Demetriou2001} models which are compared to the experimental data for the even-odd $^{153,155}$Sm isotopes, in Fig.~\ref{NLD_talys}. 

The same feature of a lower NLD for $^{155}$Sm is seen to be predicted by both microscopic models. Also shown in Fig.~\ref{NLD_talys} are the energy-dependent level density parameters $a_{stat}$ obtained within the microscopic statistical approach \cite{Demetriou2001}. The $a$ parameter is a measure of the single-particle level density at the Fermi surface and is consequently sensitive to shell and pairing effects. The difference between $^{153}$Sm and $^{155}$Sm NLDs is essentially due to stronger shell plus pairing effects in $^{155}$Sm in comparison with $^{153}$Sm leading to a smaller single-particle level density in $^{155}$Sm at the Fermi energy. This structure effect is also found in the ground-state microscopic energy predicted by most macroscopic-microscopic mass models \cite{Goriely2000,Moller2016} which give a maximum microscopic energy within the Sm neighboring isotopes for $A=148-150$ and a lower value for $^{155}$Sm compared to $^{153}$Sm.

The $^{153,155}$Sm $\gamma$SFs show pronounced strength, which is localized at mean excitation energies of about $\omega_{SR}\approx3$ MeV, a feature observed in most deformed rare-earth nuclei. The analytical technique used to extract the $\gamma$SFs, the Oslo method \cite{Schiller2000,Larsen2011}, cannot reveal fine structures in the $\gamma$-spectra. However, the observed resonances at $\sim$ 3 MeV are believed to be due to the $M$1 SR, which is consistent with other observations in this mass region. High resolution measurements (see Ref.\ \cite{Heyde2010} and references therein) have shown that these resonances are due to $M$1 transitions between high-$j$ orbitals and dominated by $J^{\pi} = 1^{+}$ states when excited directly from a $J^{\pi}=0^+$ ground state \cite{Richter1990}. The increase in the integrated SR strength from $^{153}$Sm to $^{155}$Sm is not entirely consistent with previous experimental findings and theoretical descriptions \cite{Heyde2010}. Considering that the deformation is comparable for the two nuclei the strengths should be comparable as well. A possible explanation might be that the reported calculated deformation for $^{155}$Sm is underestimated and therefore an experiment to measure both $^{153,155}$Sm deformation is highly desirable. 

%A possible explanation follows from the decrease in the NLD from $^{153}$Sm to $^{155}$Sm and concurrently the increase in the SR strength. This could suggest that the number of eligible final states becomes fewer in $^{155}$Sm (lower NLD) and therefore the strength becomes less distributed, leading to a more pronounced resonance. 

In contrast to what was observed for the actinides \cite{Guttormsen2012,Tornyi2014,Laplace2016}, the SR in the odd-even rare-earth $^{153,155}$Sm $\gamma$SF does not exhibit any double-hump structure.

\begin{figure}
\centering
\includegraphics[scale=0.45]{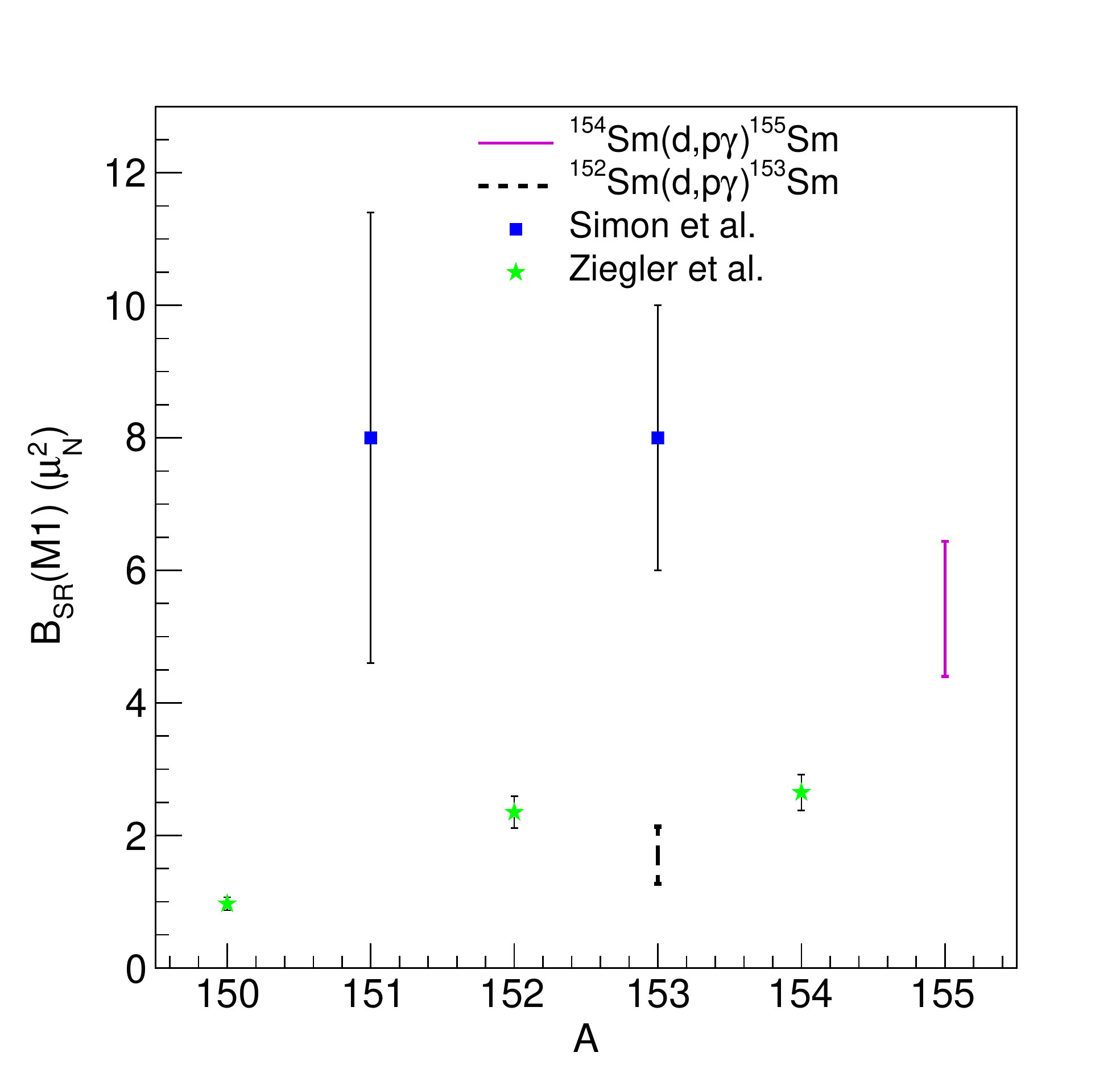}
\caption{(Color online) The experimental low-lying $M1$ strength, $B_{SR}(M1)$ plotted against mass number $A$. The present measurements (lines) extracted over a full energy range are compared to $^{151,153}$Sm data from Ref.\ \cite{Simon2016} (blue squares) and $^{150,152,154}$Sm from NRF measurements of Ref.\ \cite{Ziegler1990} (green stars). }
\label{BM1_155}
\end{figure}

Fig.\ \ref{BM1_155} compares the present integrated $B_{SR}(M1)$ with the experimental NRF values and those of Ref.\ \cite{Simon2016} extracted with the Oslo method analysis. Given the comparable deformation parameters in Tab.\ \ref{ResPar}, the $B_{SR}(M1)$ for (d,p$\gamma)^{155}$Sm is in reasonable agreement with the measurements of Ziegler \textit{et al.} \cite{Ziegler1990} on $^{152,154}$Sm when extracted over the same energy region, while the $B_{SR}(M1)$ for (d,p$\gamma)^{153}$Sm is lower than that of the neighbouring nuclei. Also the $B_{SR}(M1)$ for (d,p$\gamma)^{155}$Sm is in excellent agreement with measurements of the other rare-earth nuclei extracted with the Oslo Method \cite{Voinov2001,Melby2001,Siem2002,Guttormsen2003,Agvaanluvsan2004,Nyhus2010,Simon2016,Renstrom2018}. The results for $^{151,153}$Sm \cite{Simon2016} (also listed in Tab.\ \ref{ResPar}) are significantly higher than those from both the $^{150,152,154}$Sm NRF values and the present measurements. 

%Given the comparable deformation parameters in Tab.\ \ref{ResPar}, the $B_{SR}(M1)$ for (d,p$\gamma)^{153}$Sm is in reasonable agreement with the measurements of Ziegler \textit{et al} \cite{Ziegler1990} on $^{152,154}$Sm, while the $B_{SR}(M1)$ for (d,p$\gamma)^{155}$Sm is slightly stronger than that of the neighbouring nuclei. The results for $^{151,153}$Sm \cite{Simon2016} (also listed in Tab.\ \ref{ResPar}) are significantly higher than those from both the $^{150,152,154}$Sm NRF values and the (d,p$\gamma)^{153,155}$Sm measurements. 

This follows predictably from the discussion in Sec.\ (\ref{sec5}) where the SR appears much more pronounced in the $\gamma$SFs of (p,d$\gamma)^{151,153}$Sm data. To explain this, and the LEE discrepancy, it is important to highlight the differences in the analyses performed. One difference between how the two data sets were analyzed is the region of extraction of the NLDs and $\gamma$SFs from the primary $\gamma$-ray matrices, as the (p,d$\gamma)^{153}$Sm data were extracted for $E_{\gamma}$ as low as 0.645 MeV and from $E_{x}$ = 2.525 MeV to $E_{x}$ = 4.045 MeV. Depending on the statistical nature of the nucleus at these energies, the resulting $\gamma$SFs might be inconsistent. There might also be some residual transitions at low $\gamma$-energies in the primary $\gamma$-ray matrices for the (p,d$\gamma$)$^{153}$Sm data that lead to the differences in the $\gamma$SFs below 2 MeV. Importantly, the approaches for extracting the SR are different, for (p,d$\gamma)^{151, 153}$Sm a fit for the energy region E$_\gamma$ = 0 - 11 MeV is performed. Lastly, the highest data points for the $\gamma$SFs of (p,d$\gamma)^{153}$Sm have high uncertainties and are located at $\gamma$-energies which could give the appearance of a large SR and depending on the reliability of the last data points, the slopes could be in sharp contrast.

% Secondly, the different reactions used to populate the nucleus could populate different spin ranges.

In previous measurements, the $B_{SR}(M1)$ has proven to be less fragmented and stronger in even-even nuclei than in even-odd nuclei \cite{Heyde2010,Pietralla1998}. The strength seen in ($\gamma$,$\gamma$') experiment for odd-even cases can differ to a great extent. However, the unobserved strength can be estimated from a fluctuation analysis of the data, see Ref.\ \cite{Enders1997}. If this is taken into account, the same accumulated strengths as in the even-even cases and the same dependence on deformation is found. Different types of experiments and theoretical calculations that extract the $B_{SR}(M1)$ over the same energy region, yield similar strengths as shown for Dy isotopes \cite{Renstrom2018}. The present $^{155}$Sm measurement is in reasonable agreement whereas $^{153}$Sm appears to contract these findings. When the extraction performed here is limited by the same integration limits in Eq.~(\ref{B_sr1}), the $^{153}$Sm $B(M1)$ value is found to be significantly lower (higher limit of 1.9 {$\mu^{2}_{N}$}) than measurements of the other rare-earth nuclei extracted with the Oslo Method \cite{Renstrom2018} and that of the neighboring even-even deformed $^{154}$Sm isotope of Ref.\ \cite{Ziegler1990}. %The work of Ref. \cite{Kroll2013} reveals a significant $B_{SR}(M1)$ in the odd-even rare-earth Gd and Dy isotopes too and the reduced strength in even-even nuclei may be attributed to missing strength from unresolved $\gamma$-ray transitions as discussed in, e.g. Ref.~\cite{Heyde2010}.

A steep increase in the $\gamma$SF of $^{155}$Sm above $E_\gamma >4$~MeV might be an indication of a PDR. This excessive strength is also observed in the $\gamma$SF of $^{153}$Sm, as shown in Fig. \ref{155_3Sm_SR}. However, it is not possible to determine the electromagnetic nature of the resonance from Oslo method-type experiments and therefore information from other experiments such as NRF \cite{Kneissl1996} or inelastic proton scattering measurements \cite{Tamii2009,Neveling2011} is crucial. 
The LEE is not seen in the $\gamma$SFs of $^{153,155}$Sm for the energy range under investigation. This may be due to the current experimental conditions, which limit the extraction of useful data below $\sim$ 1 and 1.6~MeV, respectively. This is in contrast to the (d,p$\gamma$)$^{151,153}$Sm data shown in Fig.\ \ref{GSF}. 

\begin{figure}
	\centering
	\includegraphics[scale=0.42]{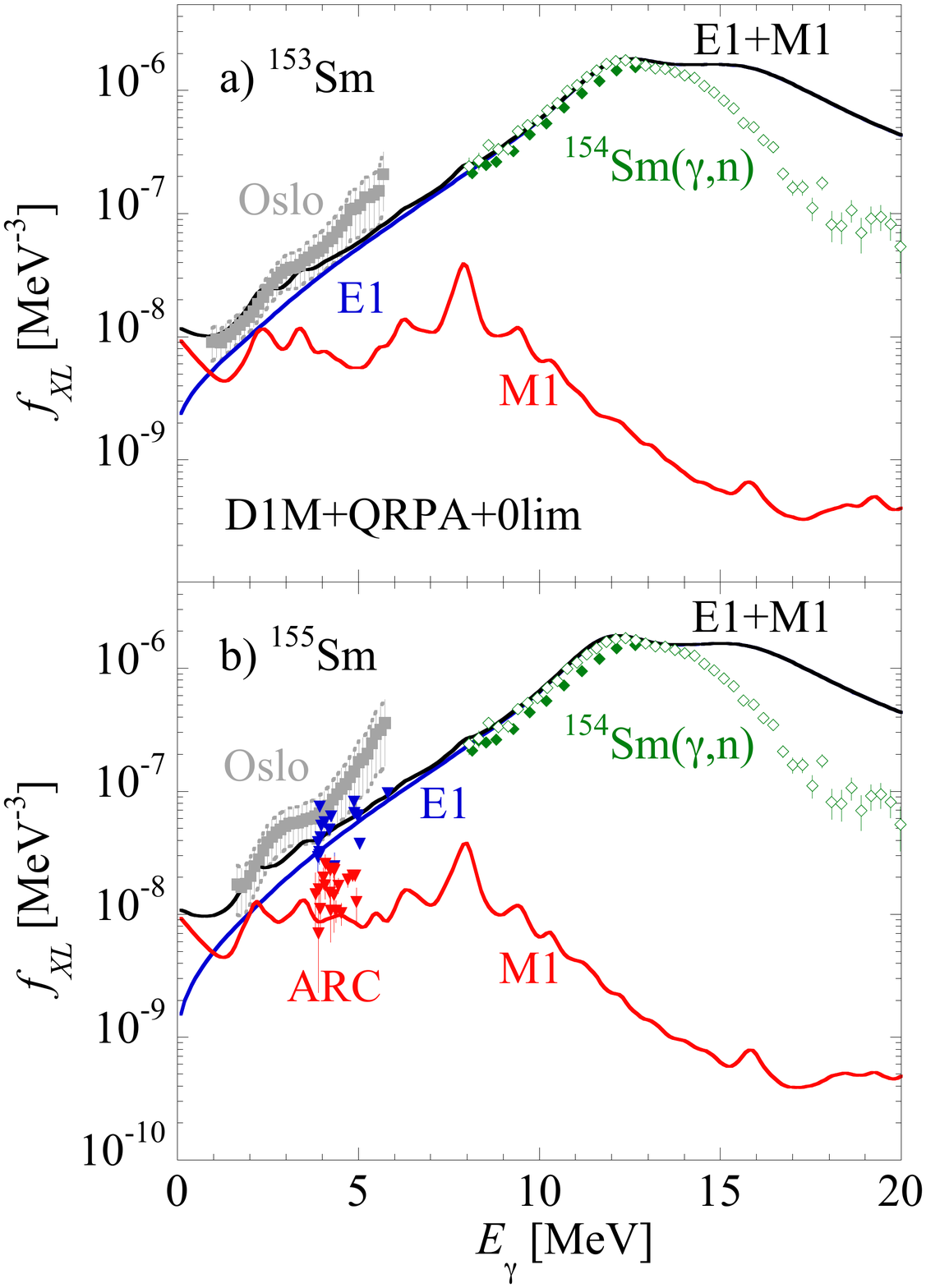}
	\caption{(Color online) Comparison between the D1M+QRPA+0lim \cite{Goriely2018a,Goriely2019} $E1$ (blue solid lines), $M1$ (red solid lines) and $E1+M1$ (black solid lines) with experimental data, a) for $^{153}$Sm and b) for $^{155}$Sm. The present Oslo data including systematic uncertainties correspond to gray squares and the ARC data to triangles (blue for $E1$ and red for $M1$). To give a fair approximation of the $\gamma$SF in the GDR region, the $\gamma$SF extracted from photoneutron cross section of the neighboring $^{154}$Sm is shown by solid \cite{Filipescu2014} and open triangles \cite{Carlos1974}. }
	\label{fig_psf_exp_d1m}
\end{figure}

Finally, we compare in Fig.~\ref{fig_psf_exp_d1m} the D1M+QRPA+0lim $E1$, $M1$ and $E1+M1$ $\gamma$SF with available experimental data, {\it i.e.} the present Oslo data, ARC data known separately for $E1$ and $M1$ strengths in $^{155}$Sm, and the $\gamma$SF extracted from photoneutron cross section of the neighboring $^{154}$Sm isotope. The D1M+QRPA+0lim model corresponds to axially deformed Gogny-HFB plus quasi-particle random phase approximation (QRPA) predictions obtained with the D1M interaction. The model has been complemented by phenomenological shell-model-inspired $E1$ and $M1$ LEE contributions to describe the de-excitation strength function \cite{Goriely2018a,Goriely2019}. As seen in Fig.~\ref{fig_psf_exp_d1m}, the D1M+QRPA+0lim predictions are in rather good agreement with data, especially for $^{153}$Sm. In the $^{155}$Sm case, the total calculated $\gamma$SF remains somewhat lower than the one found with the Oslo method though the agreement with ARC data is good. For both Sm isotopes, the large $\gamma$SF around 5~MeV cannot be explained by the D1M+QRPA+0lim model. Due to the phenomenological inclusion of an $M1$ LEE, the D1M+QRPA+0lim model can reproduce rather well the low-energy points found by the Oslo method below and around 2~MeV.

\section{\label{sec8}Summary}

The NLD and the $\gamma$SF of the deformed even-odd $^{153,155}$Sm have been measured with the reaction (d,p$\gamma$) below $B_n$ at the Oslo Cyclotron Laboratory (OCL) using the Oslo method. A pronounced resonance, the SR, was observed for both nuclei. The SR integrated strengths, in $^{153,155}$Sm were determined to be in the range 1.3\ ---\ 2.1 and 4.4\ ---\ 6.4 {$\mu^{2}_{N}$}, respectively. These values are comparable to those of neighboring nuclei for $^{155}$Sm and somewhat smaller in $^{153}$Sm.

The experimental NLD of $^{153}$Sm is found by the Oslo method to be larger than that of $^{155}$Sm, a counter-intuitive pattern that is confirmed by microscopic models and explained by stronger pairing plus shell effects in $^{155}$Sm. QRPA calculations based on the D1M Gogny interaction are also found to predict the $^{153,155}$Sm $\gamma$SF in fairly good agreement with the Oslo data, though the large strength around 5~MeV is not be described by the model. 

\section*{Acknowledgments}

The authors thank the cyclotron team at the University of Oslo for providing high-quality experimental conditions. This work is based on the research supported in part by the National Research Foundation of South Africa (Grant No.\ 118846, 92600, 90741 and 92789) and by the IAEA under Research Contract 20454. This work was partially supported by the Fonds de la Recherche Scientifique - FNRS, the Fonds Wetenschappelijk Onderzoek - Vlaanderen (FWO) under the EOS Project No O022818F and the U.S. Department of Energy under Contract DE-AC52-07NA27344. The authors gratefully acknowledge funding from the Research Council of Norway (NFR) project grant No.\ 263030 (A.G., S.S., F.Z. and V.W.I.), project grant No.\ 262952 (G.M.T.) and by the Deutsche Forschungsgemeinschaft (DFG) under Grant No. SFB 1245, project ID 279384907 (P.v.N.-C.). A.C.L. gratefully acknowledges support by the European Research Council through ERC-STG-2014 under grant agreement No. 637686.

\section*{References}
\bibliography{REf_backup}

\end{document}